\documentclass[%
 reprint,
 amsmath,amssymb,
 aps,
]{revtex4-2}

\usepackage{graphicx}% Include figure files
\usepackage{xcolor}
\usepackage{dcolumn}% Align table columns on decimal point
\usepackage{float}
\usepackage{bm}% bold math
\usepackage [english]{babel}
\usepackage{mathrsfs}
\usepackage [autostyle, english = american]{csquotes}
\MakeOuterQuote{"}
\usepackage{hyperref}% add hypertext capabilities
\newcommand{\rh}{r_{\rm h}}

\newcommand{\rhoorb}{{\rho_{\rm orb}}}
\newcommand{\Morb}{M_{\rm orb}}
\newcommand{\xihm}{\xi_{\rm hm}}
\newcommand{\rt}{r_{\rm t}}
\newcommand{\hMsun}{h^{-1}\ M_\odot}
\newcommand{\rockstar}{\texttt{Rockstar}}
\newcommand{\hMpc}{h^{-1}\ {\rm Mpc}}
\newcommand{\rhst}{r_{\rm h,st}}
\newcommand{\aacc}{a_{\rm acc}}
\newcommand{\arf}{a_{\rm RF}}

\newcommand{\tR}{{\tilde R}}
\newcommand{\avg}[1]{\langle #1 \rangle}
\newcommand{\rhmod}{r_{\rm h,mod}}

\newcommand{\rs}{r_{\rm s}}

\newcommand{\rrs}{\left( \frac{r}{\rs} \right)}
\newcommand{\rrt}{\left( \frac{r}{\rt} \right)}
\newcommand{\rsrt}{\left( \frac{\rs}{\rt} \right)}

\newcommand{\rrsa}{\rrs^\alpha}

\newcommand{\rrtb}{\rrt^\beta}

\newcommand{\rsrtb}{\rsrt^\beta}

\newcommand{\hkpc}{h^{-1}\ {\rm kpc}}

\begin{document}

\title{The Density Profile of Dynamical Halos}% Force line breaks with \\

\author{Tristen Shields}
    \email{tdshield@arizona.edu}
    \affiliation{Department of Physics, University of Arizona, Tucson, AZ 85721, USA}
\author{Edgar M. Salazar}
    \affiliation{Department of Physics, University of Arizona, Tucson, AZ 85721, USA}    
\author{Eduardo Rozo}
    \affiliation{Department of Physics, University of Arizona, Tucson, AZ 85721, USA}    
\author{Aakanksha Adya}
    \affiliation{Department of Physics, University of Arizona, Tucson, AZ 85721, USA}
\author{Calvin Osinga}
    \affiliation{Department of Astronomy, University of Maryland, College Park, MD 20742, USA}
\author{Ze'ev Vladimir}
    \affiliation{Department of Astronomy, University of Maryland, College Park, MD 20742, USA} 
%\author{Nikhil Garuda}
%    \affiliation{Department of Physics, University of Arizona, Tucson, AZ 85721, USA}

\date{\today}% It is always \today, today,

\begin{abstract}
Among the most fundamental properties of a dark matter halo is its density profile. Motivated by the recent proposal by \citet{garcia_etal23} to define a \it dynamical halo \rm as the collection of orbiting particles in a gravitationally bound structure, we characterize the mean and scatter of the orbiting profile of dynamical halos as a function of their orbiting mass. We demonstrate that the orbiting profile of individual halos at fixed mass depends on a single dynamical variable --- the halo radius $\rh$ --- which characterizes the spatial extent of the profile. The scatter in halo radius at fixed orbiting mass is $\approx 16\%$. Only a small fraction of this scatter arises due to differences in halo formation time, with late-forming halos being more compact (smaller halo radii). Accounting for this additional correlation results in an $\approx 11\%$ scatter in halo radius at fixed mass and halo formation time.
\end{abstract}
\maketitle

\section{Introduction}

One of the most basic properties of a dark matter halo is its density profile. It is typically characterized by simple analytic expressions --- e.g. the standard Navarro, Frenk, and White \citep{nfw} or Einasto profiles \citep{einasto65} --- which are fit to the total mass distribution in a halo interior to some reference radius, e.g. $R_{\rm 200m}$ (the radius enclosing a density equal to 200 times that of the mean matter density of the Universe). However, it is now broadly recognized that the dynamical structure of halos extends well past this radius. E.g., the \it splashback radius \rm \citep{diemerkravtsov14,adhikari2014,moreetal15,mansfield_etal17}, which extends well beyond $R_{\rm 200m}$, has been proposed as a more physically-motivated definition of the spatial extent of a halo \citep[see also][]{garciaetal20,fonghan21}. However, the splashback radius is flawed as a definition of the material content of the halo: on scales comparable to the splashback radius, most of the mass has never experienced a single pericentric passage. This material is not orbiting within the halo; it is falling into the halo \citep{diemer22}. These orbiting and infalling components occupy distinct regions of phase space \citep{garcia_etal23}. A single coarse-grained halo component cannot represent both populations without absorbing coherent infall into effective profile parameters, thereby mixing internal halo structure with ongoing accretion. Moreover, the orbiting particles are also formally gravitationally bound, while the infalling particles aren't \citep{richardson_etal26}. In short, there is a clear convergence of dynamical criteria for what constitutes a halo: a halo is comprised of its orbiting particles. Following \citep{garcia_etal23}, we will refer to halos defined in this way as \textit{dynamical halos}. By adopting this definition, we guarantee that the properties of halos are not impacted by the dynamics of the infall distribution through definitional artifacts (and vice-versa).

Adopting physically-motivated halo definitions matters. First, definitions that mix dynamical components introduce avoidable modeling artifacts. As a specific example, all discussion of evolution in halo relations have to contend with halo pseudo-evolution, a purely definitional artifact \citep{diemer_etal13}. Similarly, modeling the one-to-two halo transition with better than $\sim 10\%$ precision had long been impossible. This obstacle was again definitional: by adopting a physically motivated halo definition, \citep{salazar_etal25} demonstrated they could model the one-to-two halo transition with percent level precision using simple parametric models. A third  and final example is assembly bias, which is dominated by the impact of splashback structures that fall outside traditional halo boundary definitions. \citep{mansfield_kravstov20}. The lesson is clear: describing a physical system is simplest if our definitions respect the underlying dynamical structure of the system. 

Studies of halo structure suffer definitional artifacts as well. For instance, late forming halos have strong infall streams, while early forming halos will have little accretion. If halos are defined as spherical overdensities, then the dynamical structure of the orbiting mass of the halos at fixed spherical overdensity mass will be different, since the late forming halo will have few orbiting particles, and the early forming halo will have few infalling particles. That is, this correlation is definitionally induced. A much more interesting question is: if two halos have the same orbiting mass, is the structure of that orbiting mass correlated with the amplitude of the infall stream? 

Motivated by this discussion, we have set out to characterize the orbiting profiles of dynamical halos. The present paper supplies the missing calibration of their structure properties: at fixed orbiting mass, we determine which structural degrees of freedom remain, how large the intrinsic scatter is, and which correlations are genuine properties of the orbiting component rather than consequences of mixing orbiting and infalling material.

Our work is related to recent work by \citet{diemer23,diemer25}, who characterized the orbiting profile of traditional (as opposed to dynamical) halos; and the work of \citet{salazar_etal25}, who characterized the halo--mass correlation function of dynamical halos. While the two works differ on the precise parametric form used to describe the orbiting profile of halos, both of these works demonstrate the orbiting profile of halos is finite in spatial extent, with a slowly varying radial slope towards the halo center. We have chosen to use the parameterization of \citet{salazar_etal25} as our fiducial parameterization because it has fewer free parameters than that of \citet{diemer23}. We comment on how this choice impacts our results in Section~\ref{sec:diemer}.

We emphasize that although the orbiting density profile cannot be measured directly from observations, it plays an essential role in forward modeling. Previous work \citep{salazar_etal25,diemer22,diemer23} has shown that fitting a dynamics-based model for the combined orbiting and infall components yields percent-level accuracy in the observed total density profile and correctly recovers the underlying dynamical structure. Calibrating the orbiting profile is therefore valuable both theoretically and for improving dynamical interpretations of mass distributions inferred from lensing and clustering.

Our paper fills an important gap in the current literature. \citet{salazar_etal24} studied the \textit{average} orbiting profile of halos, but did not characterize halo-to-halo variations or their dependence on accretion history. By contrast, \citet{diemer22} and \citet{diemer23} studied how the orbiting profiles of halos depend on accretion history, but did so using traditional halo definitions. Our work is the first to characterize the accretion-history dependence of orbiting profiles for dynamically defined halos.

Our paper is organized as follows: in Section~\ref{sec:sim}, we describe our simulation data set, and how the particles near halos are labeled as either orbiting or infalling. We also define an accretion time for each particle, which we will use later in the paper. In Section~\ref{sec:orb_model}, we describe the orbiting profile from \citet{salazar_etal25}, and present our fits to the orbiting profiles of individual halos. We further demonstrate how the two parameters governing the halo profile are strongly correlated, enabling us to characterize individual halos profile using a single free parameter, the halo radius $\rh$. We also characterize the scatter in halo radius at fixed orbiting mass. In Section~\ref{sec:halo_radius_formation}, we investigate how the halo radius correlated with a halo's accretion history, and construct a model for the halo radius that incorporates our findings. In Section \ref{sec:diemer}, we compare our model to that of \citet{diemer23}. In Section~\ref{sec:summary} we summarize our results and discuss their implications. 

\section{Simulation and Halo Catalog}
\label{sec:sim}

This paper relies on the Cold Dark Matter (CDM) simulation described in \cite{Banerjee_sim}. This simulation was run using GADGET-2 \cite{springel_2005} and uses a 1 $h^{-1}$ Gpc box with 1024$^3$ particles. The simulation's softening length is 15 $h^{-1}$ kpc, and its cosmological parameters are $\Omega_m = 0.3$, $\Omega_{\Lambda} = 0.70$, $n_s = 0.96$, $h = 0.7$, $A_s = 2 \times 10^{-9}$, and $\sigma_s = 0.85$. The initial conditions are set at $z = 99$ using N-GenIC \cite{springel_2015}. Here, we focus exclusively on halo density profiles at $z$ = 0. To generate a dynamical halo catalog, we start with the catalog generated using the \rockstar\ halo finding algorithm \cite{behroozietal13}. Halo masses in the catalog are defined using a standard spherical overdensity definition with a threshold of 200 with respect to the mean. 

With the \rockstar\ halo catalog in hand, we proceed to find the orbiting particles for every halo using the algorithm of \citet{garcia_etal23}. Briefly, we start from the most massive \rockstar\ halo. Particles within $5\ \hMpc$ of the halo are tagged with the time at which the particle first crosses the scale associated with the one-to-two halo transition, defined as the minimum in $r^2\xihm$ \citep[][]{garcia_etal23}. We refer to this time at the accretion time, which we denote $\aacc$. Particles with early accretion times and approximately zero time-averaged radial velocities are tagged as ``orbiting,'' while particles with late accretion times and negative time-averaged radial velocities are tagged as ``infalling.'' The halo is then redefined as the collection of all orbiting particles and assigned an orbiting mass $\Morb$, the total mass of the orbiting particles. These orbiting particles are then removed from consideration when selecting orbiting particles for all subsequent halos, ensuring each particle orbits at most one halo. We then move on to the next largest halo, find its orbiting particles, and iterate until we run through the original halo catalog. The end result is a catalog of dynamical halos, with each halo having an associated list of orbiting particles. 

Throughout this study, we will explore how halo properties depend on mass by binning halos in narrow mass bins. Specifically, all binned analyses below assume 9 halo mass bins with edges $\log(M_{\rm orb}) \in$~[13.40,
13.55, 13.70, 13.85, 14.00, 14.15, 14.30, 14.45, 14.65, 15.00], where the orbiting mass $\Morb$ is measured in $\hMsun$. The particle mass in the simulation is $7.75 \times 10^{10}$ $h^{-1} M_{\odot}$.

\section{The Profile of Dynamical Halos}
\label{sec:orb_model}

\subsection{Parametric Fits}
\label{sec:model_fits}

We describe the orbiting profile of dynamical halos using the parametric model of \citet{salazar_etal25}, given by
\begin{equation}
    \label{eq:orb_model}
    \rho_{\rm orb}(x|M_{\rm orb}) = A \left( \frac{x}{\epsilon} \right)^{-\alpha(x | M_{\rm orb})}\exp\left( -\frac{x^2}{2} \right).
\end{equation}
Here, $x \equiv r / r_{\rm h}$, where $r_{\rm h}$ is the length scale at which the exponential truncation of the orbiting profile ``turns on.''  Following \citet{salazar_etal25}, we refer to this scale as the \it halo radius. \rm $A$ is a normalization constant chosen such that $\rho_{\rm orb}(x)$ integrates to the halo's orbiting mass, and as such is not a parameter in the model (we never fit for $A$). Lastly, $\epsilon$ is a mass-independent scaling parameter governing the run of the slope of the density profile $\alpha$. Specifically, $\alpha(x)$ takes the functional form
\begin{equation}
    \label{eq:slope}
    \alpha(x|M_{\rm orb}) = \alpha_{\infty} \frac{x}{x + \epsilon}.
\end{equation}
Note the slope smoothly transitions from $\alpha = 0$ to $\alpha = \alpha_{\infty}$ at scales $r \gg \epsilon r_{\rm h}$. The model has a total of three parameters: $\rh$, $\alpha_\infty$, and $\epsilon$. We follow \citet{salazar_etal25} in setting $\epsilon=0.037$, leaving only two free parameters with which to fit the orbiting profile of dynamical halos. \citet{salazar_etal25} fit the \it stacked \rm density profile of dynamical halos, finding both $\alpha_\infty$ and $\rh$ vary with mass as power-laws,
\begin{eqnarray}
    r_{\rm h, st}(M_{\rm orb}) & = & r_{\rm h, p}\left( \frac{M_{\rm orb}}{M_{\rm p}} \right)^{r_{\rm h, s}}, \label{eq:rh_power_law}\\
    \alpha_{\infty, \rm st}(M_{\rm orb}) & = & \alpha_{\rm \infty, p}\left( \frac{M_{\rm orb}}{M_{\rm p}} \right)^{\alpha_{\rm \infty, s}}. \label{eq:alpha_inf_power_law}
\end{eqnarray}
The subscripts `p' and `s' denote the pivots and slopes, respectively, while the subscript `st' indicates that these parameters are measured using halo stacks of fixed mass. The pivot mass is $M_{\rm p}=10^{14} \hMsun$. As expected, as halos become more massive they also become more extended and less concentrated (i.e. $\alpha_\infty$ decreases). Table~\ref{tab:ml_salazar} summarizes the best fit values for these parameters as reported in \citet{salazar_etal25}.

\begin{table}
    \centering
    \caption{Maximum likelihood parameter values from \citet{salazar_etal25} from Equations (1)-(4).}
    \begin{tabular}{|r|l|c|c|c}\hline
        Parameter & Description & Value\\\hline 
        $r_{{\rm h}, p}$ & Halo radius pivot (kpc h$^{-1}$) & $840.3 \pm 2.2$ \\
        $r_{{\rm h}, s}$ & Halo radius power & $0.226   \pm 0.003$ \\
        $\alpha_{\infty, p}$ & Asymptotic slope pivot & $2.018 \pm 0.002$ \\
        $\alpha_{\infty, s}$ & Asymptotic slope power & $-0.050 \pm 0.001$ \\
        $\epsilon$ & Inner scaling & $0.037 \pm 0.001$ \\
        $M_{\rm p}$ & Pivot mass & $10^{14} M_{\odot}$ \\ \hline
    \end{tabular}
    \label{tab:ml_salazar}
\end{table}

\begin{figure*}
    \centering
    \includegraphics[width=1.0\linewidth]{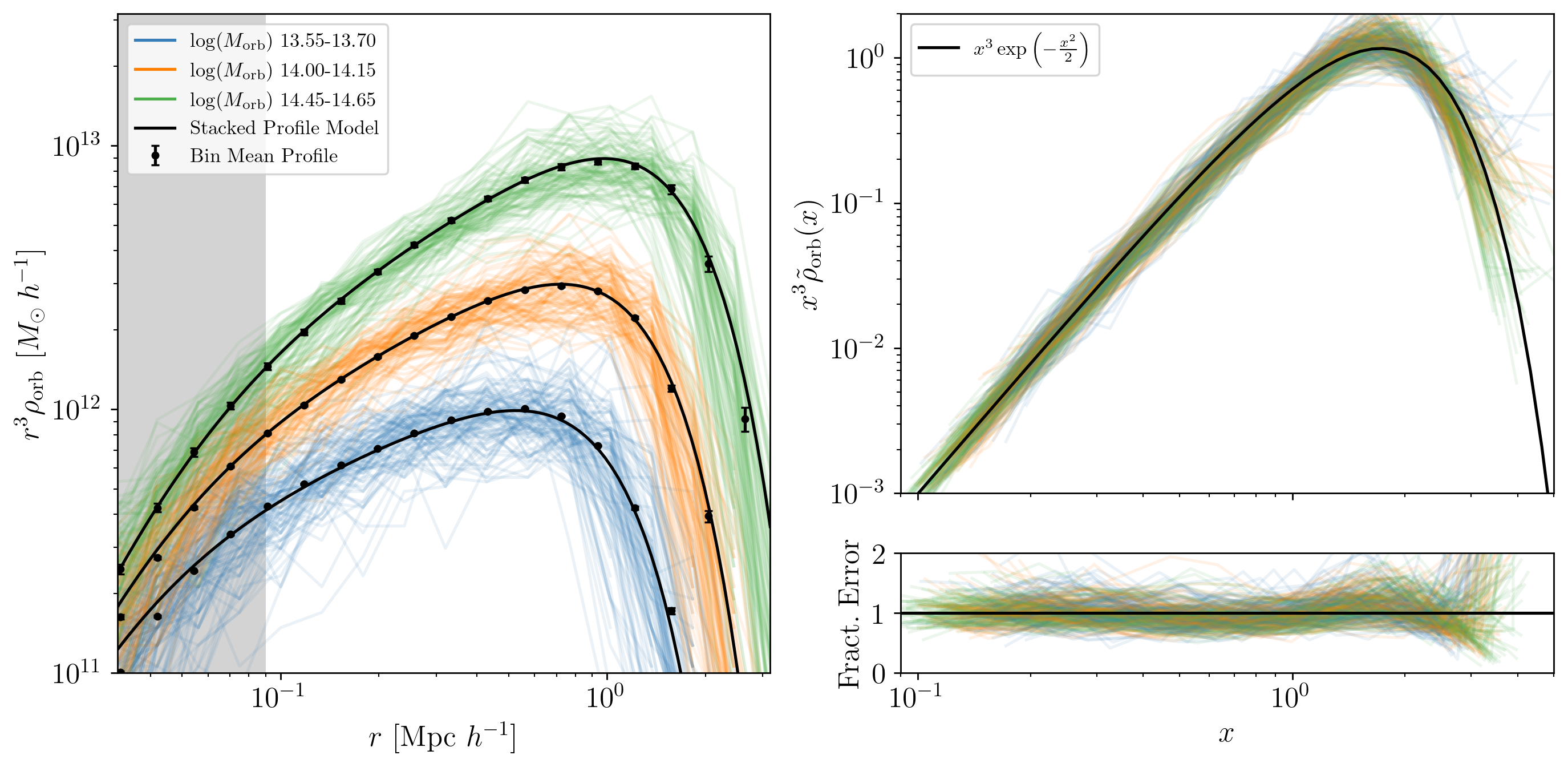}
    \caption{\textbf{Left}: Orbiting density profiles ($r^3\rho_{\rm orb}$ vs. $r$) of 300 random halos in three mass bins as labeled. Each mass bin has 100 randomly selected halos. Also shown as points with error bars are the mean orbiting profiles from all halos in the bin. The solid black lines are the best fit model obtained using Equation~\ref{eq:orb_model}. Scales $r\leq 0.09\ \hMpc$ (gray band) are subject to force-softening, and are therefore not fit. \textbf{Top-Right}: Scaled density profiles $x^3\tilde{\rho}_{\rm orb}$ (see Equation~\eqref{eq:rho_tilde}) of the same 300 halos shown at left, as a function of $x\equiv r/\rh$, where $\rh$ is the best-fit halo radius for each individual halo. The black solid line is the expectation for the rescaled profiles, $x^3\exp(-x^2/2)$. {\bf Bottom-Right:} Fractional error between the profiles from individual halos and each halo's own best fit model.}
    \label{fig:model_fitting}
\end{figure*}

The left panel in Figure~\ref{fig:model_fitting} shows the orbiting density profiles for 100 random halos in three different mass bins, as labeled. Also shown as black solid curves are the best fit parametric model of \citet{salazar_etal25} to the \it stacked \rm orbiting profile. Motivated by the excellent agreement between the model stacked profile and the individual orbiting profiles, we proceed to fit each individual halo with Equation~\eqref{eq:orb_model}. When doing so, we do not include points within the gray shaded area, corresponding to radii smaller than six times the softening length. We adopt a cost function in which the error term is the sum in quadrature of a Poisson uncertainty in the number of particles at any given radius, and a fixed per-cent error in the number of particles. 
This results in the cost function
\begin{equation}
    \label{eq:chi_squared}
    \chi^2 = \sum_{i} \frac{\left(N_{{\rm data}, i} - N_{\rm orb}(r_{i})\right)^2}{(\delta N_{{\rm data}, i})^2 + N_{{\rm data}, i} + 1}.
\end{equation}
Here, $N_{\rm orb}$ is the number of orbiting particles predicted at any given radius. Following \citet{diemer22}, we set $\delta=0.05$, corresponding to a 5\% error in the counts. We have verified our results are only marginally sensitive to the precise value of $\delta$ chosen. Finally, we add a ``+1'' term in the variance to ensure the variance is non-zero when $N_{\rm data}=0$. With this cost function in hand, we find the best fit halo radius $\rh$ and slope $\alpha_\infty$ for every halo in the catalog. 

To illustrate the quality of our fits, for each halo we calculated the scaled density profile $\tilde\rho_{\rm orb}$ defined via 
\begin{equation}
    \label{eq:rho_tilde}
    \tilde{\rho}_{\rm orb}(x) \equiv \frac{1}{A}\left( \frac{x}{\epsilon} \right)^{\alpha(x)}\rho_{\rm orb}(x).
\end{equation}
In the above Equation, $\rho_{\rm orb}$ is the density profile measured in the data as a function of the scaled radius $r/\rh$. When scaled in this way, the resulting orbiting profile is expected to be independent of halo mass, with
\begin{equation}
    \tilde\rho_{\rm orb}(x) = \exp(-\tfrac{1}{2}x^2).
\end{equation}
This scaling illustrates that the exponential cutoff of the profile is controlled simply by the scaling radius $r_h$. The right panel in Figure~\ref{fig:model_fitting} compares the scaled density profile $\tilde\rho_{\rm orb}$ for the same 300 halos shown in the left panel. The solid black line is the corresponding expectation $\tilde\rho_{\rm orb}=\exp(-\tfrac{1}{2}x^2)$ (note we actually plot $x^3 {\tilde\rho}_{\rm orb}$). The bottom--right panel corresponds to the relative difference between the scaled profiles and our model expectation. As we can see, the residuals scatter around zero. 

We compare the quality of our fits to the orbiting halo profiles to that of an NFW fit to the total profile. To do so, we fit our halos with the traditional NFW profile and compare the scatter over the radial range $r / R_{200\rm m} \in [0.3, 0.8]$. For the scatter in orbiting profiles, we measure the residuals using the model from Eq.~\ref{eq:orb_model} and the orbiting density in the simulation. For the scatter in the NFW profiles, we measure the residuals using the total density in the simulation. We then measure the median residual, and the mean absolute deviation (MAD) of both residuals in the same low, medium, and high mass bins shown in Figure \ref{fig:model_fitting}, and take $1.4826\times{\rm MAD}$ as the scatter in residuals. For the low, medium, and high mass bins, the typical median residual is $\approx -0.02$ for both our fits to the orbiting profile and for NFW fits to the total profile. The corresponding scatter in the residual is $\sim$0.15 for both fits. This demonstrates the ability to fit the dynamically-defined orbiting component of halos with residuals similar to that of NFW profiles applied to traditionally-defined halos.

\subsection{On the Degrees of Freedom of Orbiting Profiles}
\label{sec:halo_radius}

Having fit every halo with Equation~\eqref{eq:orb_model}, we wish to characterize how halos can vary in spatial extent ($\rh$) and slope $(\alpha_\infty)$ at fixed mass. To that end, we first isolate how the profile of an individual halo differs from the corresponding stacked profile by defining
\begin{eqnarray}
    \label{eq:delta_alpha}
    \Delta \alpha_{\infty} & \equiv & \alpha_{\infty} - \alpha_{\infty, \rm st}(M_{\rm orb}) \\
    \label{eq:R}
    R & \equiv & \frac{r_{\rm h}}{r_{\rm h, st}(M_{\rm orb})}
\end{eqnarray}
where $r_{\rm h,st}(\Morb)$ and $\alpha_{\rm \infty,st}(\Morb)$ are the stacked halo radius and slope given by Equations~\ref{eq:rh_power_law} and \ref{eq:alpha_inf_power_law}. 

\begin{figure}
    \centering
    \includegraphics[width=1.0\linewidth]{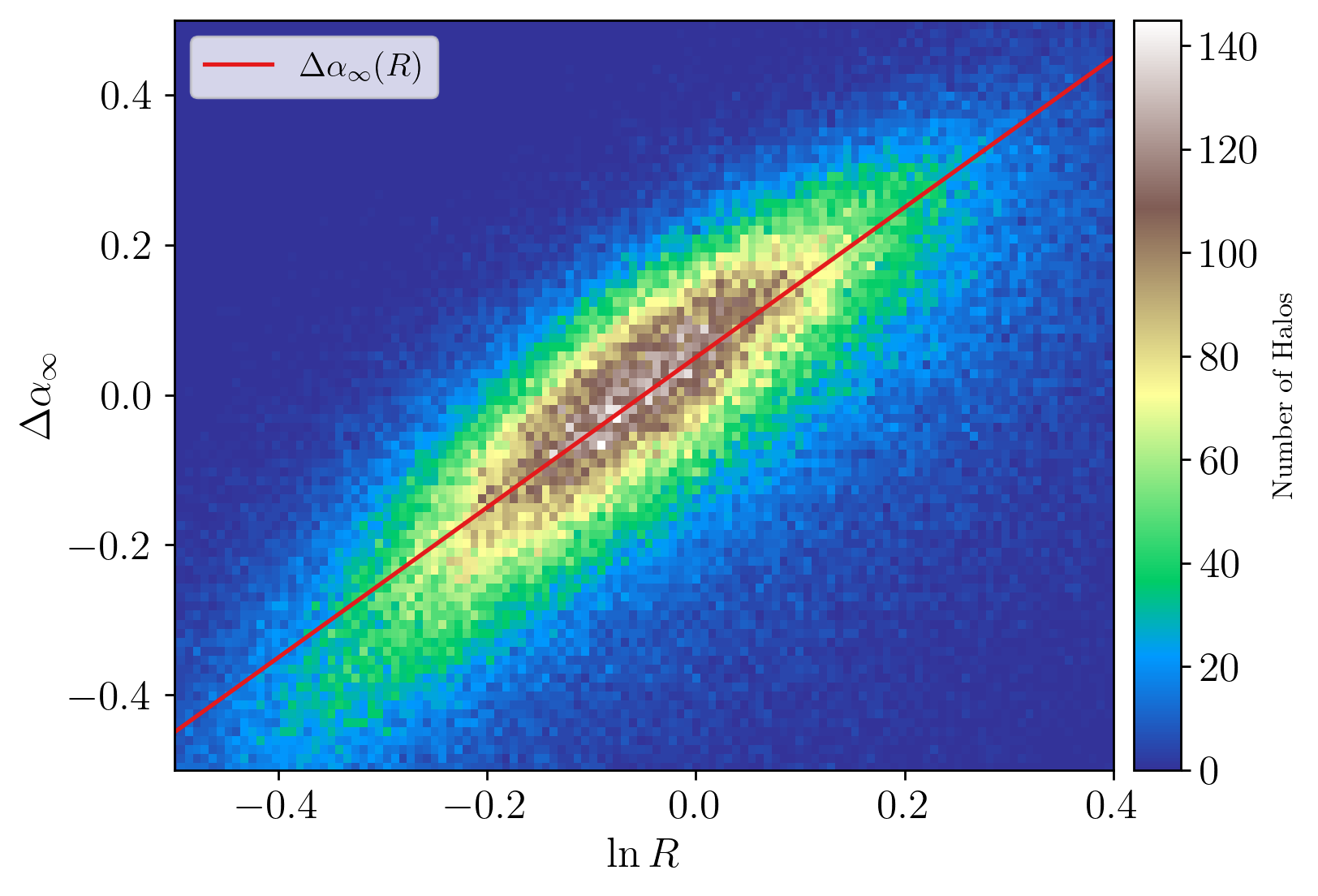}
    \caption{The color scale shows the number of dark matter halos in each pixel in the space of slope offset $\Delta\alpha_{\infty}$ (Equation~\ref{eq:delta_alpha}) and rescaled halo radius (Equation~\ref{eq:R}). We see the inner slope of a halo is tightly correlated with the scaled halo radius, a correlation we summarized by our best-fit line shown in red.}
    \label{fig:alpha_inf_calibration}
\end{figure}

In Figure~\ref{fig:alpha_inf_calibration} we show how simulated halos populate the $\Delta\alpha_\infty$--$R$ space. Evidently, the two parameters are strongly correlated, so that at fixed mass, more extended halos (higher $\rh$) having steeper inner slopes. The red line is obtained by maximizing a cost function given by the number of haloes with a $\Delta \alpha_{\infty}$ of $\pm 0.03$ of the line. The resulting best-fit line is given by  
\begin{equation}
    \Delta\alpha_{\infty} = (0.049 \pm 0.002) + (1.002\pm 0.021) \ln R
\end{equation}
where the error bars come from Jackknifing the box.
In light of these values, we have opted to adopt
\begin{equation}
    \Delta\alpha_{\infty} = 0.05 + \ln R
    \label{eq:alpha_inf_rh_relation}
\end{equation}
as our fiducial model. From these results it is clear that while our parametric fits contains two free parameters, the tight correlation between $\alpha_{\infty}$ and $\rh$ implies that halo profiles have a single effective degree of freedom, namely the halo radius $\rh$. In principle, we could choose either of the parameters to be our single degree of freedom, but $r_h$ is a natural choice since it governs the physical size of the halo's orbiting profile. A halo of mass $\Morb$ and radius $\rh$ has a density slope
\begin{equation}
    \alpha_\infty = \alpha_{\rm \infty,st}(\Morb) + \ln[\rh/\rhst(\Morb)] +0.05. \label{eq:alpha_model}
\end{equation}

We emphasize that the existence of a single radial scale is not equivalent to the statement that NFW profiles are a one-parameter family. In the traditional approach, the single parameter is the concentration, and the outer profile is set by the chosen halo boundary. In contrast, previous work on the orbiting component itself \citep[e.g.][]{diemer22,diemer23} has argued that an accurate description requires two distinct spatial scales: a scale radius $r_s$ and a truncation radius $r_t$. Our analysis of dynamical halos shows that these two structural degrees of freedom collapse onto a tight one-parameter family, which in our case allows to to find $\alpha_\infty$ given $M_{\rm orb}$ and $\rh$. That is, the apparent need for a second scale is largely a matter of modeling choice rather than an intrinsic physical degree of freedom of the orbiting profile.

\subsection{The Distribution of Halo Radii at Fixed Mass}
\label{sec:rhscat}

\begin{figure}
    \centering
    \includegraphics[width=1\linewidth]{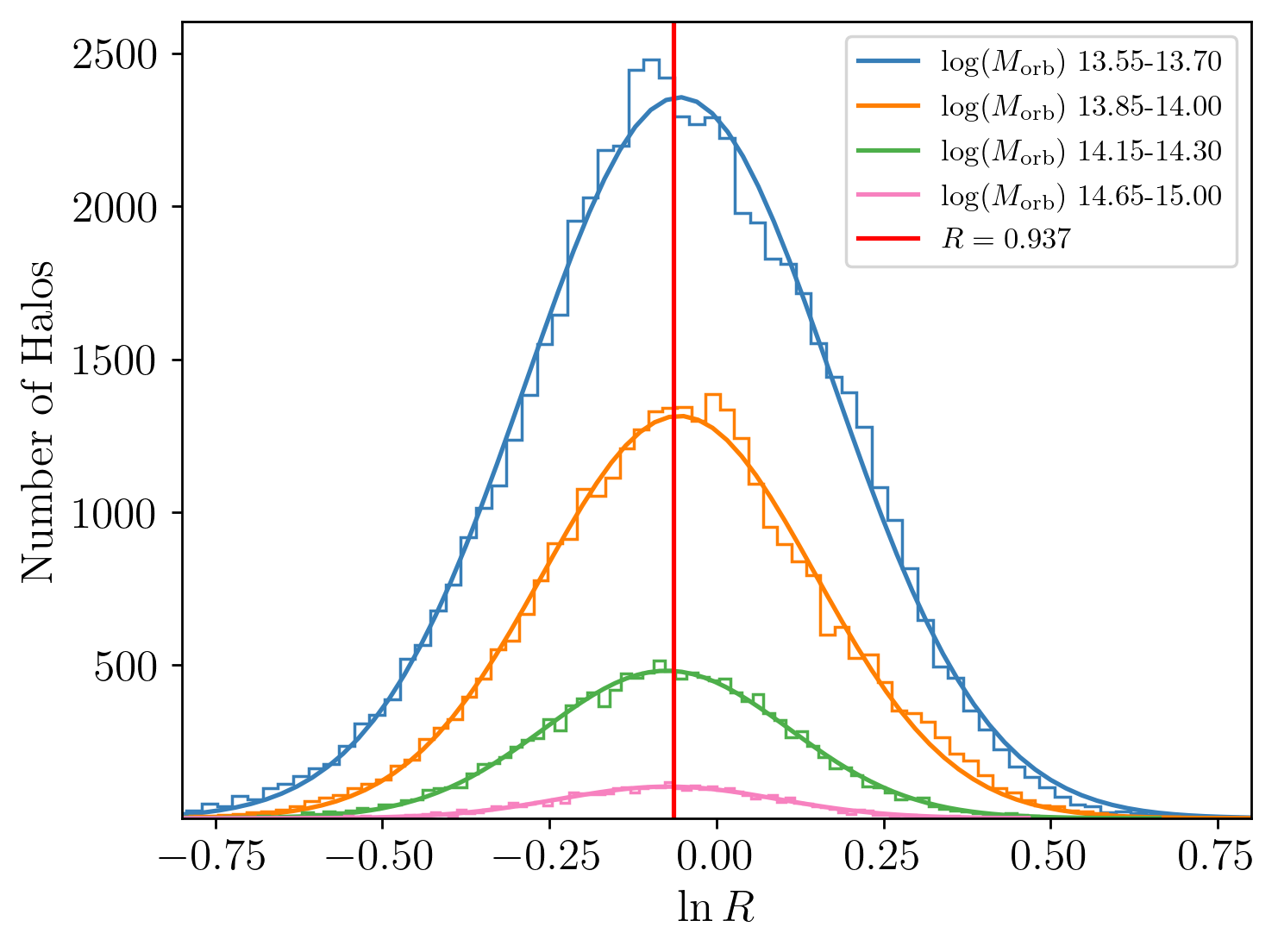}
    \caption{The distribution of $\ln \rh/\rhst$ in bins of fixed mass. We see that the halo radius $\rh$ is roughly lognormally distributed, with an average value $\avg{\ln \rh/\rhst|\Morb}=\ln 0.937 = -0.064$. After accounting for particle noise, the intrinsic scatter in $\ln \rh/\rhst$ at fixed mass is $\sigma_{\ln \rh|\Morb}=0.16$.}
    \label{fig:rhdist}
\end{figure}

In light of the above results, we have refit every halo using the halo radius as the only free parameter in the model, with the slope $\alpha_\infty(\rh,\Morb)$ set as per Equation~\eqref{eq:alpha_model}. The quality of the fits is very similar to that illustrated in Figure~\ref{fig:model_fitting}, and is therefore not shown. Figure~\ref{fig:rhdist} shows that the distribution of log-radii $\ln \rh/\rhst(\Morb)$ in bins of mass is approximately Gaussian. All distributions peak at the same ratio 
\begin{equation}
    \frac{\rh}{\rhst}=0.937,
\end{equation}
with the width of the distributions increasing with decreasing mass. We attribute this mass dependence to increasing measurement uncertainties as the number of particles in a halo decreases. Indeed, we find that we can accurately describe the mass dependence of the variance using a fit of the form
\begin{equation}
    {\rm Var}(\ln \rh/\rhst|N) = \sigma_0^2 + \frac{k}{N}
\end{equation}
where $N$ is the number of orbiting particles in a halo of mass $\Morb$. We adopt our best fit value $\sigma_{\ln \rh|\Morb}= 0.16$ as our best fit value for the scatter in $\rh$ at fixed mass. 

\section{Dependence of the Halo Radius on Accretion History}
\label{sec:halo_radius_formation}

\begin{figure*}
    \centering
    \includegraphics[width=1\linewidth]{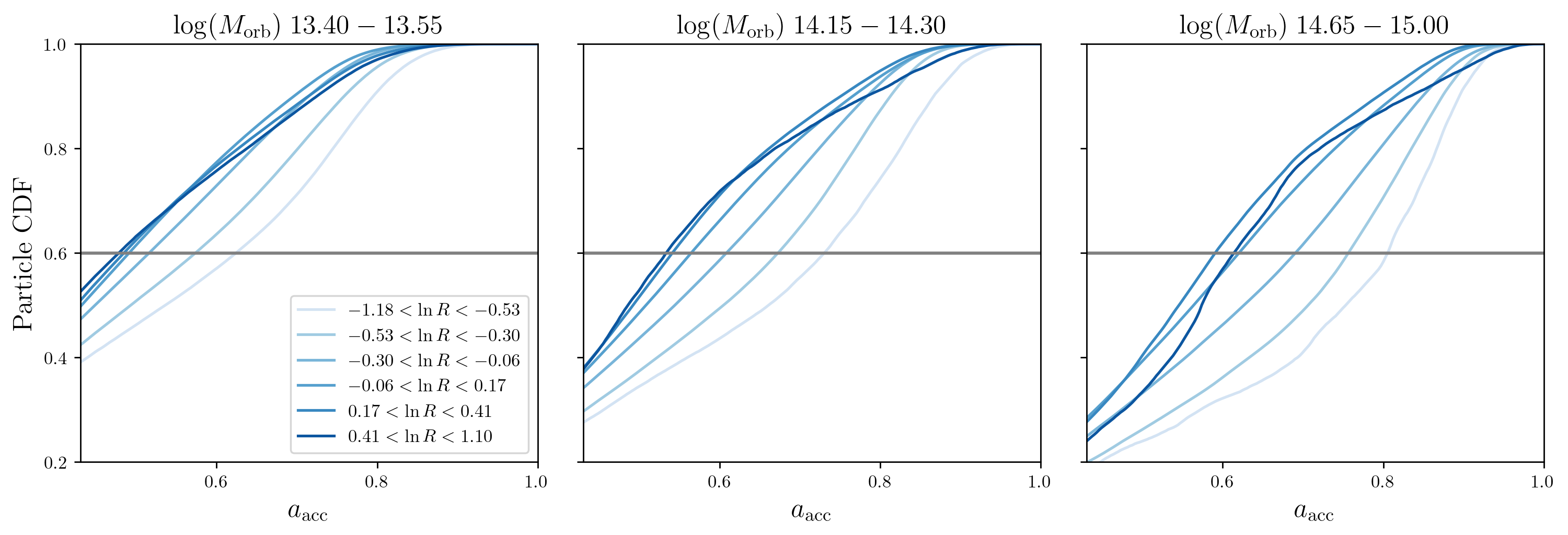}
    \caption{After binning all of the halos by $M_{\rm orb}$, then binning again by $\ln R$, we take CDFs of the particle accretion times of all the particles belonging to halos in those bins. A horizontal line is drawn when the CDF reaches 0.6, as at this point there is visually a clear distinction between most of the $\ln R$ bins that is approximately constant with mass.}
    \label{fig:a_acc_CDFs}
\end{figure*}

It has long been known that the structure of a halo is correlated with its formation history \citep[e.g.][]{wechsler_etal02}. We wish to test the extent to which the orbiting profile of a halo can be predicted using its accretion history, as characterized by the distribution of accretion times $\aacc$ of the orbiting particles in a halo. To that end, we look for differences in the distribution of particle accretion times for extended (large $\rh$) and compact (small $\rh$) halos at fixed mass. Figure ~\ref{fig:a_acc_CDFs} shows the Cumulative Distribution Functions (CDF) of particle accretion times ($\aacc$) for halos split by the scaled halo radius $R\equiv \rh/\rhst$. Each mass bin is split into six logarithmically spaced $R$ bins, with each bin being one standard deviation in $\ln R$ wide. Because $\aacc$ was measured only up to $a=0.4$, particles with earlier accretion times are assigned an accretion time of $\aacc=0.4$. This means our CDFs are correct for $\aacc>0.4$, but are unresolved below this threshold. 

The fact that massive halos form at late times leads to a systematic rightwards shift in the CDF for $\aacc$ as one moves towards more massive halos. Of more interest to us, however, is how the CDF of halos depend on $\rh$ at fixed halo mass. It is obvious from Figure~\ref{fig:a_acc_CDFs} that halos with different halo radii also have different accretion histories: the most compact halos form at late times, while the most extended halos form at early times. 

To quantify the differences in accretion history for halos of different halo radii, we wish to define a parameter that cleanly separates the different CDF curves within each mass bin. Upon inspection, we define $a_{60}$ as the value of $\aacc$ at which the CDF reaches
\begin{equation}
    \label{a_60_defn}
    \textup{CDF}(a_{60}) = 0.60
\end{equation}
and use it to differentiate early halos from late forming halos. Unfortunately, the fact that more massive halos form later implies that $a_{60}$ correlates with halo mass. This is undesirable for the purposes of characterizing \it relative \rm differences in formation. Consequently, we define the \it relative formation time \rm variable $\arf$ via
\begin{equation}
    \label{eq:a_RF_defn}
    a_{\rm RF} \equiv \frac{a_{60}}{{\rm median}(a_{60}|\Morb)}.
\end{equation}
Note that since the earliest particle accretion time in our dataset is $\aacc=0.4$, the relative formation time of halos with $a_{60}<0.4$ is not resolved. For this reason, for the remainder of this analysis we focus exclusively on halos with $\Morb>10^{14}\ \hMsun$. At these high masses, nearly all ($>99\%$) halos have $a_{60}>0.4$. 

\begin{figure*}
    \centering
    \includegraphics[width=0.47\linewidth]{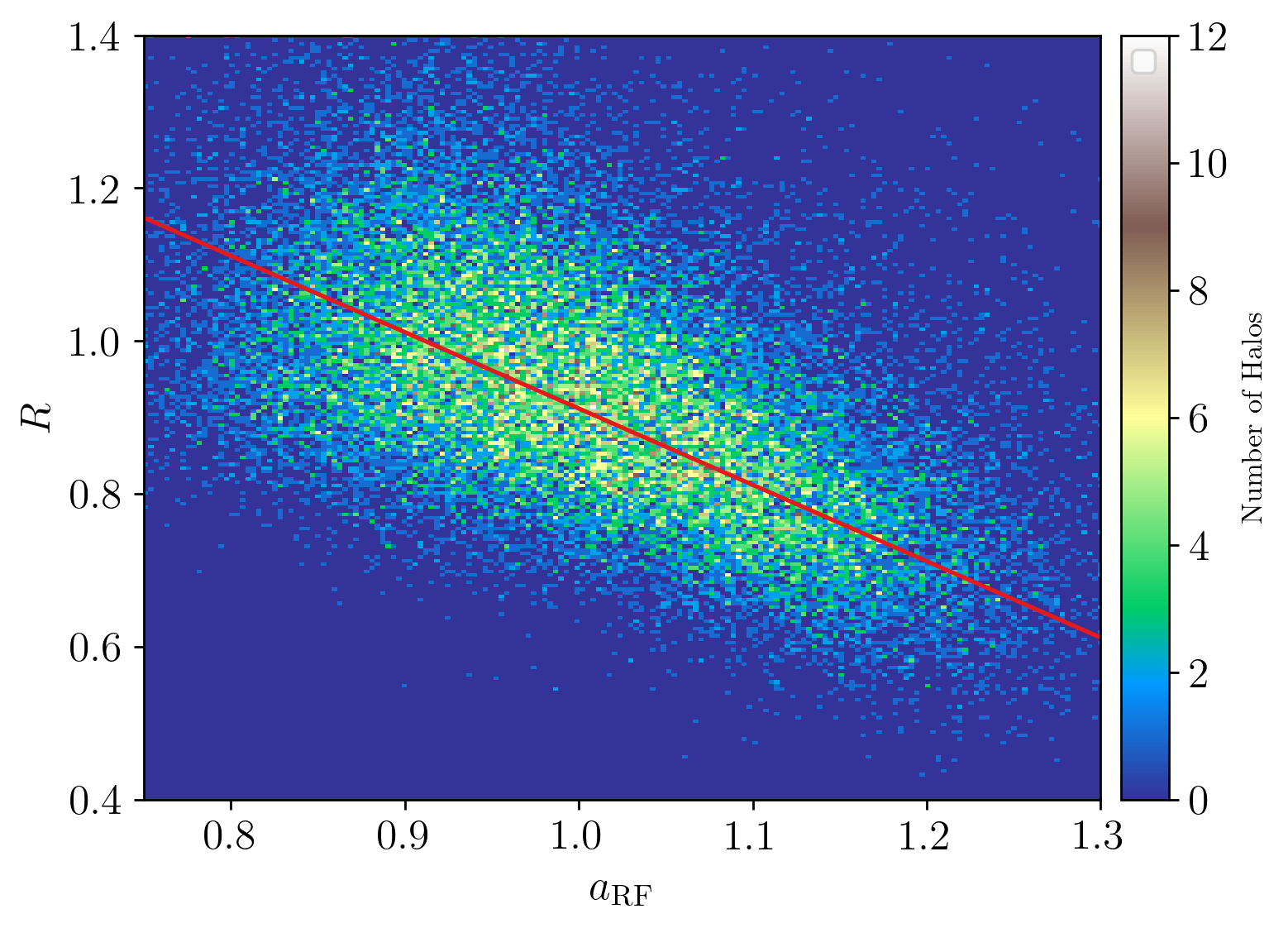} \hspace{12pt}
    \includegraphics[width=0.47\linewidth]{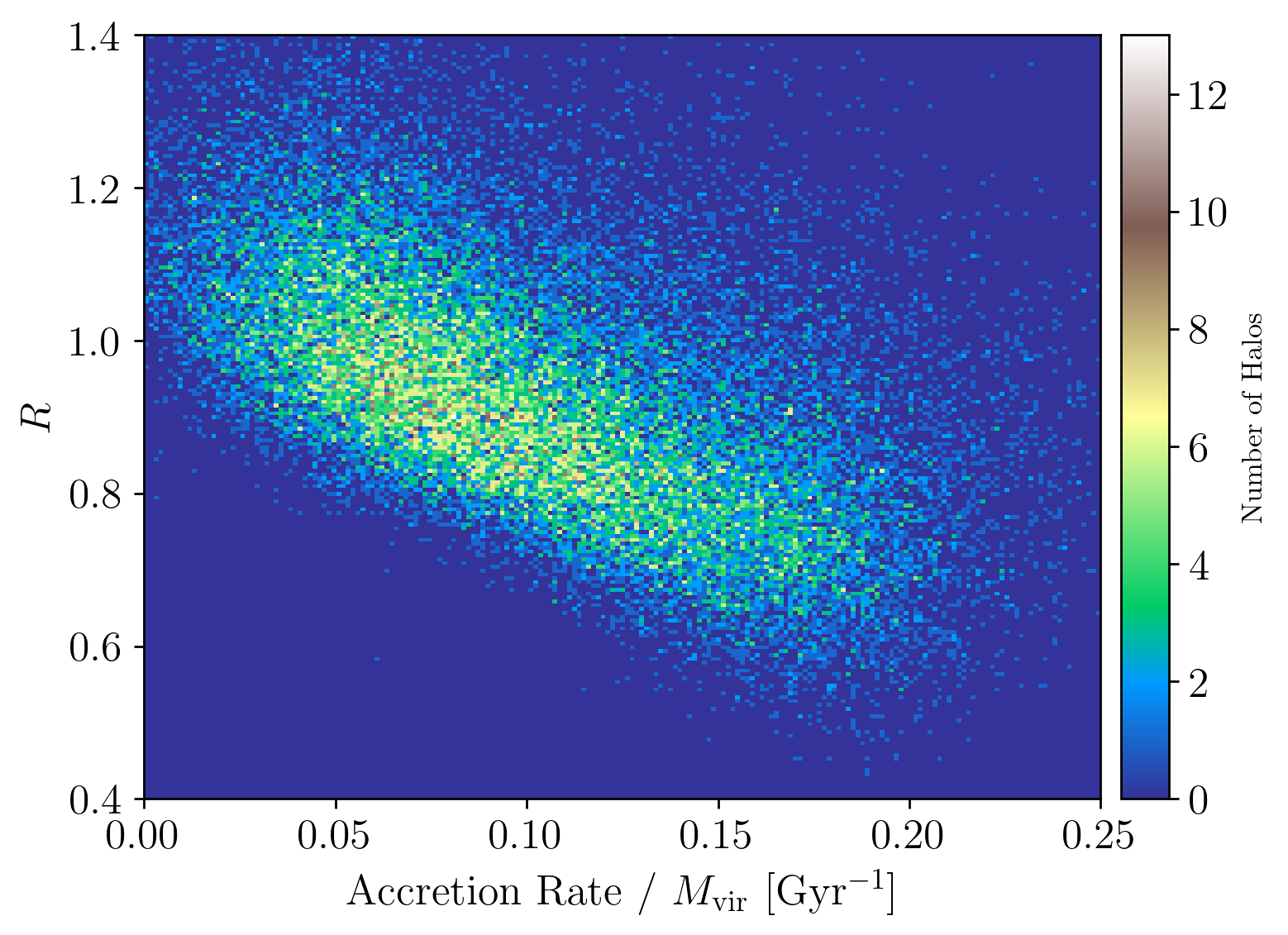}
    \caption{{\bf Left:} Correlation between the normalized halo radius (Equation~\ref{eq:R}) and a halo's relative formation time (Equation~\ref{eq:a_RF_defn}) at fixed mass, for halos of mass $M_{\rm orb} > 10^{14} \hMsun$. Our best fit line for this correlation is shown in red. {\bf Right:} Correlation between halo radius at fixed halo mass, as a function of halo accretion rate divided by halo virial mass. The qualitative trends is similar to that of the left panel, though the relation with accretion rate exhibits larger tails.}
    \label{fig:R_aRF_relation}
\end{figure*}

In the left panel of Figure~\ref{fig:R_aRF_relation} we plot the relative formation time of every massive ($\Morb\geq 10^{14}\ \hMsun$) halo against its corresponding halo radius. The two are clearly correlated, with
\begin{equation}
    \label{eq:R_aRF_relation}
    R = a_0 + s_a a_{\rm RF}
\end{equation}
where $a_0=1.89\pm 0.03$ and $s_a=-0.98\pm 0.02$. Errors come from jackknifing the box. In light of this result, we adopt a model with $s_a=-1.0$, with a corresponding best fit value $a_0=1.91 \pm 0.01$. Thus, our expression for the halo radius takes the form
\begin{equation}
    \rhmod(a_{\rm RF}, M_{\rm orb}) = (1.91-\arf) \rhst(\Morb)
    \label{eq:rhmodel}
\end{equation}
where the subscript `mod' starts for ``model.''

Note massive halos with small halo radii both form later and have shallower profiles, consistent with the usual relation that late forming halos are less concentrated \citep[][]{wechsler_etal02}.

\begin{figure*}
    \centering
    \includegraphics[width=1\linewidth]{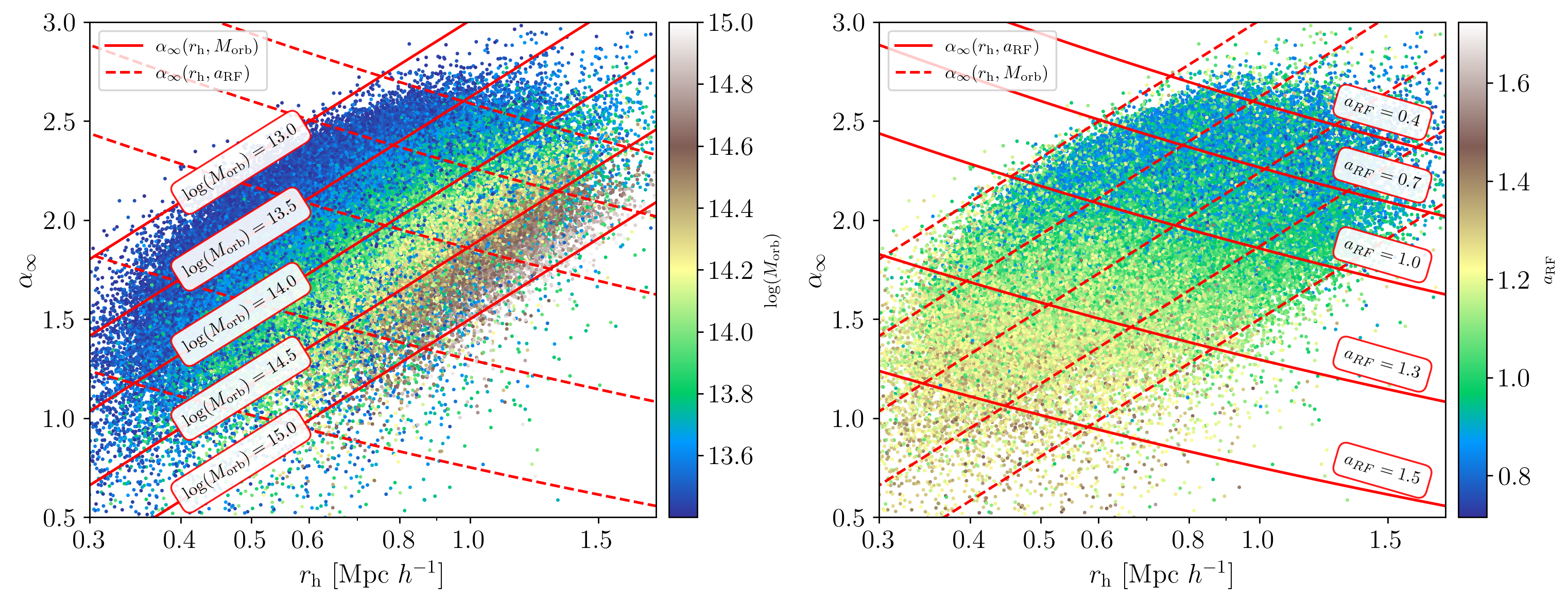}
    \caption{Visual representation of the relationships between $r_{\rm h}$, $\alpha_{\infty}$, $M_{\rm orb}$, and $a_{\rm RF}$. Both panels show how halos populate the space of halo radius $\rh$ and slope $\alpha_\infty$. In the left plot, halos are color-coded by mass, while on the right they are color coded by relative formation time. The solid red lines in the left plot show contours of fixed mass in Equation~\ref{eq:alpha_inf_rh_relation}, as labeled. Changing $\arf$ moves a halo along these lines. The solid lines in the right hand show contours of $\arf$ as labeled. The solid lines on the left panel are shown as dotted lines on the right panel, and vice versa. Varying mass at fixed relative formation time moves halos along these lines.}
    \label{fig:rh_alpha_relations}
\end{figure*}

Unfortunately, the relation between halo formation time and halo radius in Figure~\ref{fig:R_aRF_relation} is noisy. Consequently, there remains significant scatter in the distribution of halo radius at fixed mass and halo formation time. In particular, as in Section~\ref{sec:rhscat}, we find the distribution of
\begin{equation}
    \tR \equiv \frac{\rh}{\rhmod(\Morb,\arf)}
\end{equation}
is lognormally distributed with $\avg{\ln \tR|\Morb}=0.018 \pm 0.003$, independent of mass. As in Section~\ref{sec:rhscat}, the scatter is mass dependent due to particle noise; the best fit intrinsic scatter in $\rh$ at fixed mass and halo formation time is $\sigma_{\ln \rh|\Morb,\arf}= 0.11$. This is down from the 0.16 scatter in radius at fixed mass in the absence of any information about a halo's formation time, but only marginally so. Evidently, differences in the relative formation time parameter are insufficient to explain the scatter in halo radius at fixed mass. For reference, we can also consider the best fit halo radius $\rh$ obtained when fitting the density profiles with $\alpha_\infty$ as a free parameter, and then using Figure~\ref{fig:rh_alpha_relations} to predict $\rh$ as a function of $\Morb$ and $\alpha_\infty$. When we do so, the distribution of halo radii has a core with a scatter of only $\approx 5\%$, though large non-gaussian tails appear. That is, relative to the predictions based on relative formation time, there is significant room for improvement with regards to our ability to predict $\rh$ from halo properties.

Before moving on, we test whether our results on halo radii are internally self-consistent. Specifically, we have found both that
\begin{equation}
    \avg{\ln \rh/\rhmod|\Morb,\arf} = 0.018
    \label{eq:avg1}
\end{equation}
and
\begin{equation}
    \avg{\ln \rh/\rhst|\Morb} = -0.064.
    \label{eq:avgrhmorb}
\end{equation}
These two values must be related to each other. Multiplying Equation~\ref{eq:avg1} by $P(\arf|\Morb)$ and integrating, we find after some algebra that
\begin{equation}
    \avg{\ln (\rh/\rhst)|\Morb} = 0.018 + \avg{\ln(1.91-\arf)|\Morb}. 
    \label{eq:rhpred}
\end{equation}
Evaluating the right hand side of the above Equation by averaging over the distribution of relative formation times we obtain $\avg{\ln (\rh/\rhst)|\Morb}=-0.094$. This value differs from our best fit by $0.029$. We attribute this $3\%$ difference to the impact of non-Gaussian tails in the distribution of halo radii.

\section{Comparison to Literature}
\label{sec:diemer}

While there is an extensive literature on the properties of the
\textit{total} matter density profile, our results are not directly
comparable to them \citep[e.g.][]{nfw,einasto65} because our fits are for the orbiting component alone. Nevertheless, the
\textit{quality} of the fits can be compared. As noted before (Sec.~\ref{sec:model_fits}), we find that the mean
and rms residuals of an NFW fit to the total density profile (fit out
to $2R_{200}$) are comparable to those of our orbiting--profile fits.
Moreover, standard results for total--profile fits---such as the
correlation between concentration and formation time
\citep{wechsler_etal02}---are qualitatively consistent with the trends we
find for the orbiting component: early--forming halos are more compact
and have steeper inner slopes. 

A more direct comparison is possible with \citet{diemer25}, who analyzed the orbiting profiles of halos using a methodology similar to ours. Our two works differ in that:
\begin{enumerate}
    \item Diemer focuses on the properties of $\rhoorb$ for spherical overdensity halos, whereas we focus specifically on dynamical halos. Both works describe orbiting profiles, but the physical halo definitions differ, as we redefine our halos using the algorithm described in Section \ref{sec:sim}. Even though the difference in algorithms has a minimal impact on the orbiting profiles of individual halos \citep{vladimir_etal25}, the choice of definition matters with regards to the scatter and correlation of halo properties with mass.
    \item Diemer relies on the parametric model of \citet{diemer23}, which takes the form
\begin{equation}
    \rhoorb(r) =\rho_s \exp(S(r))
\end{equation}
where
\begin{eqnarray}
    S(r) & \equiv & -\frac{2}{\alpha} \left[ \rrsa - 1 \right] -\frac{1}{\beta} \left[ \rrtb - \rsrtb \right] .
\end{eqnarray}
\end{enumerate}
We note that while the Diemer model has four degrees of freedom ($r_t$, $r_s$, $\alpha$, and $\beta$), when fitting individual halo profiles Diemer holds the parameters $\alpha$ and $\beta$ fixed to their best-fit values in the stacked profile, $\alpha=0.18$ and $\beta=3.0$. This reduces the number of degrees of freedom in the individual fits to two, the same number of degrees of freedom in the \citet{salazar_etal25} ($\rh$ and $\alpha_\infty$). However, this reduction in parameter space is not unique, a point that we return to below. For now, we simply note that when fitting individual halo profiles, Diemer varies only the radius $r_t$ and the scale radius $r_s$: the radius $\rt$ is roughly equivalent to the halo radius $\rh$ in \citet{salazar_etal25}, while the scale radius $r_s$ is roughly equivalent to $\alpha_\infty$, as they both characterize the slope of the density profile in the core of the halo. With this understanding in hand, we show below that the two works are in agreement with regards to the qualitative trends in the profiles. Unfortunately, quantitative comparisons are non-trivial because the two works adopt different halo mass definitions. 

We begin by considering how the extent of the orbiting profile correlates with mass. \citet{diemer25} finds the ratio $\rt/R_{\rm 200m}$ decreases with increasing mass, implying the slope of $\rt$ with $M_{\rm 200m}$ is smaller than 1/3. Since $M_{\rm 200m} \approx \Morb$ \citep[Fig.~9 in][]{garcia_etal23}, this is consistent with the $\approx 1/4$ power-law index in the relation between $\rh$ and $\Morb$ in \citet{salazar_etal25}. That is, \it the spatial extent of orbiting halos scales with mass with a power-law index flatter than 1/3. \rm

We then turn to the tight correlation between $\alpha_\infty$ and $\rh$ obtained from fitting individual halos (see Figure~\ref{fig:alpha_inf_calibration}). This correlation is also apparent in the correlation between $\rs$ and $\rt$ in Figure~10 of \citet{diemer25}: in units of $R_{\rm 200m}$, halos with larger $\rt$ have a smaller $\rs$, resulting in steeper inner profiles, precisely as per Figure~\ref{fig:alpha_inf_calibration}. However, the correlation in our Figure~\ref{fig:alpha_inf_calibration} appears to have smaller scatter than the one shown in Figure~10 of \citet{diemer25}. We attribute the increased scatter in the latter work to two factors:  1) when fitting individual halo profiles, \citet{diemer25} set $\alpha=0.18$ and $\beta=3$, rather than solving for the relations $\alpha(\rt)$ and $\beta(\alpha)$ apparent in their Figure~3. That is, the reduction in parameters space in \citet{diemer25} was sub-optimal, leading to increased scatter. Second, while Figure~10 of \citet{diemer25} roughly accounts for the mass dependence of $\rs$ and $\rt$, neither of these length scales is exactly proportional to $R_{\rm 200m}$. Consequently, some of the scatter in Figure~10 of \citet{diemer25} is due to trends of $\rs/R_{\rm 200m}$ and $\rt/R_{\rm 200m}$ with mass. Despite these differences, the main conclusion is the same: \it at fixed mass, the more extended a halo is, the steeper its inner density profile is. \rm 

At this point, the comparison between our results and those of \citet{diemer25} becomes more difficult. Based on the tight correlation between $\alpha_\infty$ and $\rh$, we adopted $\rh$ as the single free parameter in our model. By contrast, \citet{diemer25} maintained both $\rs$ and $\rt$ as free parameters, albeit with significant scatter. Nevertheless, we find similar qualitative trends. \citet{diemer25} finds that while $\rt$ does not vary strongly with halo formation time, $\rs$ does: early forming halos have lower $\rs/R_{\rm 200m}$ values, resulting in steeper profiles. These early halos must therefore have large $\alpha_\infty$ values, which in turn correspond to large halo radii, i.e. \it early forming halos have larger radii, \rm in agreement with our results (see Figure~\ref{fig:R_aRF_relation}).

We have further shown that there remains significant scatter in the distribution of halo radii $\rh$ at fixed halo mass, even after accounting for the impact of the relative formation time parameter $\arf$ on $\rh$. Based on the self-similar collapse model, the splashback radius is expected to depend primarily not on halo mass, but rather on accretion rate \citep{diemerkravtsov14,shi16}. Indeed, \citet{diemer25} finds that their radial boundary $\rt$ is tightly correlated with a halo's recent accretion rate $\Gamma$ (see their Figure~12), where $\Gamma$ is the average halo mass accretion rate ($\Gamma \equiv \Delta \ln M_{\rm 200m}/\Delta \ln a$) over one dynamical time (see \citep{diemer25} for details). Similar conclusions hold for the splashback radius in simulations \citet{shin_diemer23}. 

The right panel in Figure~\ref{fig:R_aRF_relation} shows that the variation in halo radius at fixed halo mass (i.e. $R\equiv \rh/\rhst$) is correlated with accretion rate per unit mass as measured using the \texttt{Acc\_Rate\_2Tdyn} tag from {\sc ConstistentTrees} \citep{consistent_trees}. This accretion rate is the change in spherical overdensity mass of halo over a dynamical time $t_{\rm dyn}=2R_{\rm 200}/v_{\rm 200}$, divided by the present day mass. Note this definition is similar --- but not identical --- to that used in \citet{diemer25}. Comparing the left panel in Figure~\ref{fig:R_aRF_relation} to the right panel, we see that the correlation between halo radius and relative formation time is comparable to that between halo radius and accretion rate. In particular, although the latter appears to be slightly tighter, it also has more pronounced tails. Overall, we do not see a clear preference for selecting accretion rate over relative formation time as the relevant dynamical variable that controls variations in halo radii at fixed halo mass. However, we caution that we have not been able to self-consistently estimate halo accretion rates for dynamical halos, an omission we intend to address with the ability to quickly generate dynamical halo catalogs (Salazar et al., in preparation). 

\section{Summary and Implications}
\label{sec:summary}

In this work, we sought to characterize the density profile of dynamical halos. The goal is not to replace conventional density-profile models as empirical fits to the total matter distribution around individual halos. Rather, it is to calibrate the orbiting component of dynamically defined halos. By working at fixed orbiting mass and fitting only the orbiting material, the correlations measured here describe the structure of the orbiting halo itself rather than a changing mixture of orbiting and infalling material. This calibration provides the baseline needed for dynamical decompositions of halo density profiles and for interpreting which structural correlations are intrinsic to the orbiting component.

Our main results are:
\begin{itemize}
    \item The individual orbiting density profiles of dynamical halos can be fit using the $\rhoorb(r)$ parametric model of \citet{salazar_etal25} on scales $r\geq 90\ \hkpc$ with residuals comparable to that of the NFW profile for the total matter density. This model contains two free parameters: a logarithmic slope $\alpha_\infty$, and a parameter $\rh$, the exponential decay length governing the finite spatial extent of the orbiting profile. We refer to $\rh$ as the \it halo radius. \rm
    \item At fixed orbiting mass, the slope of the profile ($\alpha_\infty$) is tightly correlated with halo mass, with more extended halos having steeper slopes. This allows us to reduce the number of free parameters in the $\rhoorb(r)$ model to \it one: \rm the halo radius.
    \item At fixed mass, the halo radius is correlated with halo formation time, such that at fixed mass, halos with early forming halos having more extended profiles with steeper slopes. 
    \item The intrinsic scatter in halo radius at fixed mass is $\sigma_{\ln \rh|\Morb}=0.16$. Accounting for the correlation between halo formation time and halo radius, we are able to  reduce the scatter in halo radius to $\sigma_{\ln \rh|\Morb,\arf}=0.11$. The trends of halo radius with relative formation time are qualitative similar to those between halo radius and accretion rate.
\end{itemize}

This work is intimately related to that recently published by \citet{diemer25}, who characterized the orbiting orbiting profiles of spherical halos using the parametric model advocated for in \citep{diemer23}. While these differences make quantitative comparisons of our results difficult, we argue either model constitutes an effective parameterization of the same underlying phenomenology. We refer the reader to Section~\ref{sec:diemer} for a more comprehensive discussion of how the two works are related. 

An interesting implication of our results is that if one is able to determine the structural properties of a halo, namely its halo radius and slope $\alpha_\infty$, then one can recover both the mass and formation time of the halo. This is illustrated in Figure~\ref{fig:rh_alpha_relations}, where we plot the distribution of halos in the $(\rh,\arf)$-plane, color coded by either mass (left panel) or relative formation time (right panel). The red solid lines are contours of fixed mass (left panel) or relative formation time (right panel) calculated using Equations~\ref{eq:rhmodel} and \ref{eq:alpha_inf_rh_relation}. Evidently, one can ``read off'' the mass and formation time of a halo from its radius and slope (albeit with significant scatter). This type of physical insight is inaccessible in standard spherical-overdensity halos, where the halo radius is definitionally tied to the enclosed mass and is therefore decoupled from the underlying dynamics. Because of this imposed relation, the standard framework cannot produce an analogue of Fig.~\ref{fig:rh_alpha_relations}. By contrast, the dynamical-halo construction exposes how a halo’s extent and structure arise from its orbital dynamics, highlighting the conceptual power of the approach.

\section*{Data Availability}
\rm All simulation data is available upon request. The code used to perform this analysis is publicly available on GitHub at \url{https://github.com/tristen-shields/dark_matter_halo_profiles2025}.

%\href{https://github.com/tristen-shields/dark_matter_halo_profiles2025/tree/main}{here}.

% Need acknowledgements and paper references
\section*{Acknowledgements}

T.S. was supported in part through the Arizona NASA Space Grant Consortium, Cooperative Agreement 80NSSC20M0041. E.R. and E.S. were supported by NSF grant 2206688, and DOE grant DE-SC0009913. This research was supported in part by the National Science Foundation under Grant number 2206690. Participation of A.A. in this research was supported by the University of Arizona TIMESTEP program through the TIMESTEP Research Apprenticeship Program. E.R. gratefully acknowledges productive conversations with Andrew Zentner and Susmita Adhikari.

\def\jcap{JCAP}
\def\mnras{MNRAS}             % Monthly Notices of the RAS
\def\aap{A\&A}             % Astronomy and Astrophysics
\def\apjs{ApJS}             % Astrophysical Journal Supplement
\def\apjl{ApJL}             % Astrophysical Journal Supplement
\def\baas{BAAS}
\def\araa{ARA\&A}
\def\aj{AJ}
\def\pasj{PASJ}
\def\physrep{ßPhys.~Rep.}   % Physics Reports

\bibliography{main}% Produces the bibliography via BibTeX.

\end{document}